\documentclass[preprint,showpacs,showkeys,preprintnumbers,amsmath,amssymb]{revtex4}

\usepackage{graphicx}
\usepackage{dcolumn}
\usepackage{bm}
\begin{document}


\title{On angular momentum of gravitational radiation}

\author{Alexander I. Nesterov}
\affiliation{Departament of Physics, C.U.C.E.I., Guadalajara University,
Guadalajara, Jalisco, Mexico}
\email{nesterov@cencar.udg.mx}

\date{\today}

\begin{abstract}
The quasigroup approach to the conservation laws (Phys. Rev. {\bf
D56}, R7498 (1997)) is completed by imposing new gauge conditions for
asymptotic symmetries. Noether charge associated with an arbitrary element
of the Poincar\'e quasialgebra is free from the supertranslational
ambiquity and identically vanishes in a flat spacetime.

\end{abstract}

\pacs{PACS numbers: 04.20.Ha,04.20.-q}
\keywords{quasigroups, conservation laws, aymptotic symmetries}
\maketitle

\section{Introduction}

A major difficultes in defining of the angular momentim for isolated system
are that, for asymptotically flat spacetimes the group of asymptotic
symmetries is an infinite-parametric one (see,{\em e.g.},\cite{Held} and
references therein). The (asymptotic) symmetries group contains an unique
four-parametric translation subgroup and infinite-parametric subgroup of
supertranslations. Although the asymptotic symmetries group has a unique
translation subgroup there is no canonical Lorentz subgroup since the last
one emerges as a factor group of the asymptotic symmetries group by the
infinite dimensional subgroup of supertranslations. This leads to the main
problem in defining of angular momentum on null infinity, namely, all
existing expressions suffer from supertranslational ambiguity.

In this letter we suggest a way to resolve this problem applying the
quasigroup approach to the conservation laws in general relativity
developed in \cite{MEN,N,N1,N2,N3,N4}. In Sec. 2 we outline
the main facts from the theory of quasigroups of transformations and
consider the asymptotic symmetries group at future null infinity ($\cal
J^+$). In Sec. 3  new gauge conditions  restricting the asymptotic
symmeties group to a particulare Poincar\'e quasigroup are imposed and
compared with the twistor approach {\cite{PR1,PR2}}. In Sec. 4 starting
with the expression (linkage) given by Tamburino and Winicour \cite{TW1,W},
we define the Noether charge associated with any element of the Poincar\'e
quasialgebra and calculate the flux of the momentum and angular-momentum
emitted by a gravitational system. Concluding remarks comprise the final
Sec.  5.

We use the spin coefficients formalism by Newman and Penrose \cite{PR1}
choosing $\kappa=\varepsilon+\bar\varepsilon=0,\; \tau=\bar\pi=
\bar\alpha+\beta,\; \rho=\bar\rho$.

\section{Quasigroups of transformations and asymptotic symmetries}

Let $\frak M$ be a n-dimensional manifold and the continuos law
of transformation is given by $x'=T_a x , \quad  x\in {\frak M}$, where
\{$a^i$\} is the set of real parameters, $i=1,2,\dots,r$. The set of
transformations $\{T_a\}$ forms a $r$-parametric quasigroup of
transformations (with right action on ${\frak M}$), if \cite{Bat}:\\
1) there exists a
unit element which is common for all $x^\alpha$ and corresponds to
$a^i=0:\;T_a x|_{a=0}=x$;\\
2) the modified composition law holds:
\[
T_aT_b x
=T_{\varphi(b,a;x)} x;
\]
3) the left and right units coincide:
\[
\varphi(a,0;x)=a, \quad \varphi(0,b;x)=b;
\]
4) the modified law of associativity is satisfied:
\[
 \varphi(\varphi(a,b;x),c;x)=\varphi(a,\varphi(b,c;T_a x);x);
\]
4) the transformation inverse to $T_a$ exists: $x=T^{-1}_a x'$.

The generators of infinitesimal transformations
\[
\Gamma_i=(\partial(T_ax)^\alpha/\partial a^i)|_{a=0}
\partial/\partial x^\alpha \equiv R^\alpha_i(x)\partial/\partial x^\alpha
\]
form {\it quasialgebra} and obey the commutation relations
\begin{equation}
[\Gamma_i, \Gamma_j] =C^p_{ij}(x)\Gamma_p,
\label{eq1}
\end{equation}
where $C^p_{ij}(x)$ are the structure functions satisfying the modified
Jacobi identity
\begin{eqnarray}
C^p_{ij,\alpha}R^\alpha_k + C^p_{jk,\alpha}R^\alpha_i
+C^p_{ki,\alpha}R^\alpha_j
+ C^l_{ij}C^p_{kl} + C^l_{jk}C^p_{il} + C^l_{ki}C^p_{jl}=0 .
\label{eq2}
\end{eqnarray}

{\bf Theorem.} Let the given functions $R^\alpha_i, \;C^p_{kj}$
obey the equations (\ref{eq1}), (\ref{eq2}), then locally the quasigroup
of transformations is reconstructed as the solution of set of differential
equations:
\begin{eqnarray}
\frac{\partial\tilde x^\alpha}{\partial a^i}
= R^\alpha_j(\tilde x)\lambda^j_i(a;x),\quad
\tilde x^\alpha(0)=x^\alpha, \label{eq3} \\
\frac{\partial\lambda^i_j}{\partial a^p}
-\frac{\partial\lambda^i_p}{\partial a^j}
+ C^i_{mn}(\tilde x)\lambda^m_p\lambda^n_j=0,
\quad \lambda^i_j(0;x)=\delta^i_j \label{eq4}.
\end{eqnarray}
Eq. (\ref{eq3}) is an analog of the Lie equation, and Eq. (\ref{eq4}) is
the generalized Maurer-Cartan equation.



As it is known, for asymptotically flat at future null infinity ($\cal
J^+$) spacetime the group of asymptotic symmetries is the
infinite-parametric Newman-Unti (NU) group which contains the
infinitedimensional Bondi-Metzner-Sachs (BMS) group preserving strong
conformal geometry of $\cal J^+$ \cite{NU,HNP,BMS,S,S01}.
On $\cal J^+$ a general element of NU-algebra is given by
\begin{equation}
\xi=B(u,\zeta,\bar\zeta)\Delta^0 +
C(u,\zeta,\bar\zeta)\bar\delta^0 + \bar C(u,\zeta,\bar\zeta) \delta^0,
\label{eqG}
\end{equation}
where $\eth C=0$ and $\eth,\Delta^0 $, $\delta^0 $ are the standard NP
operators ``edth'', $\Delta$ and  $\delta$ restricted on $\cal J^+$;
$\zeta$ being a complex stereographic coordinates at two-dimensional
space-like cross sections of $\cal J^+$ labeled by a coordinate $u$ and its
metric is given by
\[
ds^2=\frac{2d\zeta d\bar\zeta}{[P(u,\zeta,\bar\zeta)]^2}.
\]
We assume $\zeta = x^2 - i x^3$ and that $P$ can be written
as
\[
P=V(u,\zeta,\bar\zeta)P_0,
\]
where $P_0=(1/\sqrt 2)(1+\zeta\bar\zeta)$ and $V$ is to be a regular
function on the sphere.  When $V=1$ the metric above is reduced to the
Bondi metric.

The generators of four-parameter translation subgroup read
\begin{equation}
\xi_a=B_a(\zeta,\bar\zeta)\Delta^0, \quad (C=0),
\end{equation}
($a$ runs from 1 to 4) with the function $B_a$ satisfying
\begin{equation}
\eth^2 B_a=B_a(4({\bar\alpha}^0)^2 - 2 \eth\bar\alpha^0),
\quad {\rm where} \quad \alpha^0 = \frac{1}{2}\bar\eth\ln P.
\label{Tr1}
\end{equation}
Its solution can be written as $B_a = l_a/V(u,\zeta,\bar\zeta)$, where
$l_a$ satisfies $\eth^2_0 l_a = 0$ and a ``standard edth'' $\eth_0$ is
referred to Bondi frame. There are four independent solutions of this
equation.

The generators of ``Lorentz group''are determined as follows:
\begin{eqnarray}
\xi_A= B_A(u,\zeta,\bar\zeta)\Delta^0
+C_A\bar\delta^0+ \bar C_A \delta^0, \quad (\eth C_A=0),
\label{Q}
\end{eqnarray}
($A$ runs from 1 to 6). Here $B_A(u,\zeta,\bar\zeta)$ is an arbitrary real
function and a dot being derivative with respect to retarded time $u$.
There are six independent solutions of the equation
$\eth C_A=0$, which can be written as
$C_A = l_A(\zeta,\bar\zeta)/V(u,\zeta,\bar\zeta)$, where $l_A$ is a
solution of the equation $\eth_0 l_A=0$.

The generators of the NU group obey the commutation relations
{\cite{N4}}:
\begin{eqnarray}
[ \xi_a ,\xi_b ]=0, \quad [\xi_a,\xi_B]=C^b_{aB}(u,\zeta,\bar\zeta)\xi_b,
\nonumber \\
\left [ \xi_A ,\xi_B \right] =  C^D_{AB}(u,\zeta,\bar\zeta)\xi_D,
\label{str}
\end{eqnarray}
where $C^b_{aB},\;C^D_{AB}$ are the {\it structure functions} depending on
an arbitrary function $B_A$. The commutation relations above show that the
NU group is in fact a {\it quasigroup} with the closed Lorentz quasialgebra
and a supertranslational ambiguity.

\section{Reduction of the NU group to the Poincar\'e quasigroup}

The key idea is to reduce the NU group to the ten-parametric quasigroup (the
Poincar\'e quasigroup), imposing the appropriate conditions on an arbitrary
function $B_A$, and thus fixing the supertranslational freedom.

The scheme consists of following \cite{N1,N3,N4}:

(i) Propagate the asymptotic generators $\xi$ inward along the null
superface $\Gamma$ intersecting $\cal J^+$ in $\Sigma^{+}$ by means of
geodesic deviation equation:
\begin{equation}
\nabla^2_l\xi + R(\xi,l)l=0,
\label{Jac}
\end{equation}
imposing the appropriate conditions at $\cal J^+$.

(ii) Use the commutation relations (\ref{str})
and asymptotic expansion of $C^a_{bc}$
\[
C= C_0 + C_{1}r^{-1} + C_{2}r^{-2} + \cdots ,
\]
$r$ being a canonical parameter and $C_0$-s the same as in Eq.(\ref{str}),
evaluate all coefficients of these series.

In the spin-coefficient notation the geodesic deviation equation
\begin{equation}
\nabla^2_l\xi + R(\xi,l)l=0,
\label{Jac_B}
\end{equation}
accomplished by substituting
\begin{equation}
\xi=AD +B \Delta + \bar C \delta + C\delta,
\end{equation}
reads
\begin{eqnarray}
D^2 A + 2(\tau D\bar C + \bar\tau DC) + CD\bar\tau + \bar C D\tau -
\nonumber \\
-  C \bar\Psi_1 - \bar C \Psi_1 - B(\Psi_2 +\bar \Psi_2) =0,
\label{sys_1}\\
D^2 B =0, \quad
D^2 C + B D\tau + \bar C \Psi_0  + B \Psi_1 =0.
\label{sys_2},
\end{eqnarray}
Using the asymptotic expansion
\begin{eqnarray}
A = A_1 r + A_0 + A_{-1}r^{-1} + 0(r^{-2}), \nonumber \\
B = B_1r + B_0 + B_{-1}r^{-1} + 0(r^{-2}), \nonumber \\
C = C_1 r + C_0 + C_{-1}r^{-1} + 0(r^{-2}), \nonumber
\end{eqnarray}
one obtains the solution of the geodesic deviation equation in the following
form:
\begin{eqnarray}
A_{-1} = \Re(B \Psi^0_2 - 2\tau^0 \bar\eth B), \quad
C_{-1} = B\tau^0, \quad
B_{-n} = 0 \;(n\geq 1).
\end{eqnarray}

The system of equations (\ref{sys_1}), (\ref{sys_2}) is the second order
ordinary differential system, and, for obtaining the unique solution, one
needs to impose initial conditions on the functions $A,B,C$ and its first
derivatives. So the only freedom in the solution is in $A_0, A_1, B_0,B_1,
C_0, C_1$. We adapt the asymptotic Killing equations for determining these
coefficients.  Starting with
\begin{eqnarray}
\lim_{r\rightarrow\infty}l^\mu
l^\nu\pounds_{\xi}g_{\mu\nu}= 0,\quad \lim_{r\rightarrow\infty}m^\mu
n^\nu\pounds_{\xi}g_{\mu\nu}= 0, \quad
\lim_{r\rightarrow\infty}m^\mu \bar
m^\nu\pounds_{\xi}g_{\mu\nu}= 0,\\
\lim_{r\rightarrow\infty}r m^\mu \bar
m^\nu\pounds_{\xi}g_{\mu\nu}= 0,\quad
\lim_{r\rightarrow\infty}rl^\mu m^\nu\pounds_{\xi}g_{\mu\nu}= 0,
\end{eqnarray}
one obtains
\begin{eqnarray}
A_0 = \eth\bar\eth B+ B\eth\bar\eth \ln P, \quad
A_1 = -\frac{1}{2}(\bar\eth C_1 + \eth \bar C_1), \quad
 B_1 = 0 \Rightarrow B=B_0, \\
C_0 = - \eth B + \sigma_0 \bar C_1,  \quad
\dot C_1 + (\ln P)\dot{} C_1 =0 \Rightarrow
C_1 = c(\zeta\bar\zeta)/V,
\end{eqnarray}
(dot being derivative with respect to retarded time $u$) and the remaining
freedom is in the functions $B, C_1$.

Now let us consider in the flat spacetime all null hypersurfaces and not
just those which are shear-free.  Then the functions $B, C_1$ must satisfy
the following equations:
\begin{eqnarray}
\eth^2  B -  B(4({\bar\alpha}^0)^2 - 2 \eth\bar\alpha^0) -\frac{\sigma^0}{2}(3 \eth\bar
G_1 - \bar \eth C_1) - \bar C_1 \eth\sigma^0 - C_1 \bar\eth\sigma^0 =0,\\
\eth C_1 =0,
\label{X}
\end{eqnarray}
($\sigma^0$ being the asymptotic shear), arising from the Killing equation
\begin{eqnarray}
m^\mu m^\nu\pounds_{\xi}g_{\mu\nu} = 0.
\label{Kil}
\end{eqnarray}
The system above is the unique one determining the
functions $B, C_1$ and restricting the NU group to a particular
Poincar\'e group. Its solution is given by
\begin{equation}
B=B_t + \Big(\eth\eta\bar C_1 + \frac{u-\eta}{2}\eth\bar C_1  + c.c.\Big )
\label{B1}
\end{equation}
where
\begin{eqnarray}
\sigma^0= \eth^2 \eta -\eta(4({\bar\alpha}^0)^2 - 2 \eth\bar\alpha^0),
\quad {\rm with} \quad \eta =u -\frac{1}{V}\int_0^u V du,
\end{eqnarray}
and $B_t$ satisfies Eq.(\ref{Tr1}):
\begin{equation}
\eth^2 B_t=B_t(4({\bar\alpha}^0)^2 - 2 \eth\bar\alpha^0),
\end{equation}

Setting $B=\xi^0 + \bar\xi^0, C_1 =\bar\xi^1$, where
\[
\xi^0=\xi^0_t + \eth\eta \xi^1 + \frac{u-\eta}{2}\eth\xi^1,
\]
we find that $\xi^0,\xi^1$ satisfy
\[
\eth^2 \xi^0 = \xi^0(4({\bar\alpha}^0)^2 - 2 \eth\bar\alpha^0)
+ \frac{1}{2}\sigma^0\eth\xi^1 + \eth(\sigma^0 \xi^1),
\quad \bar\eth\xi^1 =0.
\]

In the presence of radiation the equation (\ref{Kil}) leads to $\eth C_A=0$
and
\begin{eqnarray}
\eth^2 B - B(4({\bar\alpha}^0)^2 - 2 \eth\bar\alpha^0)
- B \bar{\cal N} =\frac{\sigma^0}{2}(3 \eth\bar G_ A - \bar \eth C_A)
+ \bar C_A \eth\sigma^0 + C_A \bar\eth\sigma^0,
\nonumber
\end{eqnarray}
where the news function ${\cal N} = \lambda^0 + 2 \bar\eth\alpha^0 -
4(\alpha^0)^2$ is introduced. In general case the solution of this equation
does not exist, because $B$ is a real function and $\sigma^0$ is the
complex function. Thus the equation above is incompatible with (\ref{B1}),
unless ${\cal N} = 0$.

A way to resolve this problem is to obtain the differential
constraints on the functions $B$ and $C_1$, which are compatible with
Eq.(\ref{Kil}) for the flat(stationary) spacetimes  and lead to the
definition of the asymptotic shear as $\sigma_0= \eth^2 \eta
-\eta(4({\bar\alpha}^0)^2 - 2 \eth\bar\alpha^0)$ for some complex function
$\eta$, so next we devote our attention to a method that fixes this
problem.

Let us consider at ${\cal J}^+$ the complex vector
\begin{equation}
\xi_c =\xi^0\Delta^0 + \xi^1 \delta^0,
\end{equation}
such that the generators (\ref{eqG}) read $\xi =\xi_c + \bar\xi_c$. This
yields $B= \xi^0 + \bar\xi^0, \; \bar C_1 = \xi^1$ and
$\bar\eth\xi^1=0$. We specify the vector $\xi_c$ as follows:
\begin{equation}
\xi^0=\xi^0_t + \eth\eta \xi^1 + \frac{u-\eta}{2}\eth\xi^1,\quad
\eth^2 \eta -\eta(4({\bar\alpha}^0)^2 - 2 \eth\bar\alpha^0) =\sigma^0,
\label{B2}
\end{equation}
where $\xi^0_t$ is the solution of the following equation:
\begin{equation}
\eth^2 \xi^0_t=\xi^0_t(4({\bar\alpha}^0)^2 - 2 \eth\bar\alpha^0),
\end{equation}
determining the translational subgroup of the NU group. The definition
(\ref{B2}) implies that $\xi^0$ satisfies
\begin{eqnarray}
\eth^2 \xi^0 = \xi^0(4({\bar\alpha}^0)^2 - 2 \eth\bar\alpha^0)
+ \frac{1}{2}\sigma^0\eth\xi^1 + \eth(\sigma^0 \xi^1),
\label{MEQ}
\end{eqnarray}
This differential equation is compatible with the Killing equations (see
discussion above) and its solution (\ref{B2}) exists in general case,
including radiation.  We adopt (\ref{MEQ}) as the master equation
restricting the NU group to a particular Poincar\'e quasigroup.

Emerging of the constraint (\ref{MEQ}) also can be
understood analyzing the twistor equation on a cross section $S$ of $\cal
J$
\begin{equation}
\nabla_{A'}{}^{(A}\omega^{B)} =0, \label{TEQ}
\end{equation}
projected on $S$ \cite{Dray,Shaw} . The components of the equation
(\ref{TEQ}) tangent to $S$ are
\begin{equation}
\bar\eth \omega^0 =0, \quad \eth\omega^1 = \sigma^0 \omega^0.
\label{TEQ1}
\end{equation}

Let $\omega^A_1, \omega^A_2$  be any two solutions of (\ref{TEQ1}). We
associate with the twistor $\omega^{AB} = \omega^{(A}_1 \omega^{B)}_2$ the
complex NU generator
\begin{equation}
\xi_c =\xi^0\Delta^0 +
\xi^1\delta^0 =-i(\omega^0_1\omega^1_2 +\omega^0_2\omega^1_1) \Delta^0
-2i\omega^0_1\omega^0_2 \delta^0.
\end{equation}
With help of (\ref{TEQ1}) one obtains that $\bar\eth\xi^1 =0$ and $\xi^0$
satisfies
\begin{eqnarray}
\eth^2 \xi^0 =
\xi^0(4({\bar\alpha}^0)^2 - 2 \eth\bar\alpha^0)
+ \frac{1}{2}\sigma^0\eth\xi^1 + \eth(\sigma^0 \xi^1),
\label{B3}
\end{eqnarray}
which is just (\ref{MEQ}). In the Bondi reference frame
(\ref{B3}) reduces to the equations obtained by Shaw (Eq.(20) in
\cite{Shaw}).

Our results can be summarized as follows:

(i) A general element of NU-algebra is given by
\begin{eqnarray}
\xi=B(u,\zeta,\bar\zeta)\Delta^0 +
C(u,\zeta,\bar\zeta)\bar\delta^0 + \bar C(u,\zeta,\bar\zeta) \delta^0,
\quad \eth C=0, \\
B=B_t + \eth\eta\bar C_1 + \frac{u-\eta}{2}\eth\bar C_1
+ \bar\eth\bar\eta C_1 + \frac{u-\bar\eta}{2}\bar\eth C_1 ,
\nonumber \\
\eth^2 \eta -\eta(4({\bar\alpha}^0)^2 - 2 \eth\bar\alpha^0) =\sigma^0,\quad
\eth^2 B_t=B_t(4({\bar\alpha}^0)^2 - 2 \eth\bar\alpha^0).
\label{eqG1}
\end{eqnarray}
(a) The generators of translations are given by
\begin{equation}
\xi_a =B_a\Delta^0,
\label{eqTr}
\end{equation}
where $B_a=l_a(\zeta,\bar\zeta)/V(u,\zeta,\bar\zeta)$, and $l_a$ satisfies
$\eth_0^2 l_a =0$.

(b) The generators of boosts and rotations are given by
\begin{eqnarray}
\xi_A =B_A\Delta^0 + C_A\bar\delta^0+ \bar C_A \delta^0,
\quad \eth C_A =0, \\
B_A = \eth\eta\bar C_A + \frac{u-\eta}{2}\eth\bar C_A + c.c., \quad
\eth^2 \eta -\eta(4({\bar\alpha}^0)^2 - 2 \eth\bar\alpha^0) =\sigma^0,
\label{eqBR}
\end{eqnarray}
where $C_A=l_A(\zeta,\bar\zeta)/V(u,\zeta,\bar\zeta)$, and $l_A$ satisfies
$\eth_0 l_A =0$.

(ii) The generators of the Poincar\'e quasigroup obey at ${\cal J}^{+}$ the
following commutation relations:
\begin{eqnarray}
[ \xi_a ,\xi_b ]=0, \quad
[\xi_a,\xi_B]=C^b_{aB}(u,\zeta,\bar\zeta)\xi_b, \nonumber \\
\left [ \xi_A ,\xi_B \right]
= C^D_{AB}(u,\zeta,\bar\zeta)\xi_D, \nonumber
\end{eqnarray}
$C^D_{AB},\;C^i_{aB}$  being the {\it structure functions}.

\section{Energy-momentum and angular momentum at $\cal J^+$}

As it is known the Komar integral \cite{K}, providing a fully satisfactory
notation of the total mass in stationary, asymptotically flat spacetimes, is
not invariant under a change of the choice of the generators of time
translations in the equivalence class associated with the given BMS
translation. For an asymptotically flat at future null infinity spacetime
the modified ``gauge invariant'' Komar integral (linkage)
\begin{eqnarray}
L_\xi(\Sigma)=-\lim_{\Sigma_\alpha\rightarrow{\cal J^{+}}}\frac{1}{4\pi}
\oint_{\Sigma_\alpha} \left( \xi^{\left [\alpha;\beta \right]} +
\xi^\rho_{;\rho} l^{\left [\alpha \right.} n^{\left .\beta \right ]}\right)
ds_{\alpha\beta},
\end{eqnarray}
where $\{\Sigma_\alpha\}$ is one-parameter family of spheres, was
introduced by Tamburino and Winicour \cite{TW1}. We adopt this as our
definition of the conserved quantities on $\cal J^+$ associated with the
generators of the Poincar\'e quasigroup. The computation leads to the
following coordinate independent expression
\begin{eqnarray}
L_\xi=-(1/4\pi)\Re \oint\left \{B\left(\Psi^0_2 + \sigma^0\lambda^0
-\eth^2\bar\sigma^0\right) \right.\nonumber \\
\left.+ 2 \bar C\left(\Psi^0_1 + \sigma^0\eth{\bar\sigma}^0
+(1/2)\eth(\sigma^0\bar\sigma^0)\right)\right\} dS,\; (\eth C =0).
\end{eqnarray}

Introducing the complex Noether charge
\begin{eqnarray}
Q_c(\xi)=-(1/4\pi) \oint\left \{\xi^0\left(\Psi^0_2 + \sigma^0\lambda^0
-\eth^2\bar\sigma^0\right) \right.\nonumber \\
\left.+ \xi^1\left(\Psi^0_1 + \sigma^0\eth{\bar\sigma}^0
+(1/2)\eth(\sigma^0\bar\sigma^0)\right)\right\} dS
\end{eqnarray}
and setting $B =\xi^0 + \bar\xi^0, \;  C_1 = \bar \xi^1$, we find
\begin{equation}
L_\xi= Q_c(\xi) + \bar Q_c(\xi).
\end{equation}
Now applying {\it Lemma 2} and {\it Lemma 3} \cite{Dray}:
\begin{eqnarray}
\oint \xi^0\left(\eth^2\bar\sigma^0 +\bigl(2\eth\bar\alpha^0 -
4(\bar\alpha^0)^2\bigr)\bar\sigma^0\right)dS
+ \oint\xi^1\left(\sigma^0\eth{\bar\sigma}^0
+\frac{1}{2}\eth(\sigma^0\bar\sigma^0)\right)dS =0,
\nonumber \\
\oint \xi^0\left(\bar\eth^2\sigma^0 +\big(2\eth\bar\alpha^0 -
4(\alpha^0)^2\big)\sigma^0\right)dS =0 \nonumber
\end{eqnarray}
together with the  Bianchi identity
$\Psi^0_2 + \sigma^0\lambda^0 -\bar\eth^2\sigma^0 = \bar\Psi^0_2 + \bar
\sigma^0\bar\lambda^0 -\eth^2\bar\sigma^0$, one can write
\begin{eqnarray}
Q_c(\xi)=-(1/4\pi) \oint\left(\xi^0\left(2\Psi^0_2 + 2\sigma^0{\cal N}
-\bar\Psi^0_2 -\bar\sigma^0\bar{\cal N}\right)
+ \xi^1\Psi^0_1\right) dS,
\label{Q1}
\end{eqnarray}
This immediatly yields $Q_c \equiv 0$ in flat spacetime.

The integral four-momentum is given by \cite{N1,N3,N4}
\begin{equation}
P_i=  -(1/4\pi)\Re \oint B_i(\Psi^0_2 + \sigma^0{\cal N}) dS,
\label{EM}
\end{equation}
where $B_i=l_i/{V}$ and four-vector
\[
l=\frac{1}{1+|\zeta|^2}\left(1+|\zeta|^2, \zeta + \bar\zeta,
i(\zeta - \bar\zeta), |\zeta|^2 - 1 \right).
\]
Using the Bianchi identities, we compute the loss of energy-momentum
\begin{equation}
\dot P_i = -(1/4\pi) \oint B_i |{\cal N}|^2 dS.
\label{P}
\end{equation}
This expression coincides with the result obtained by Lind {\it et al}
\cite{LMN} and in the Bondi frame it yields the standard formula (see, {\it
e. g.} \cite{PR1})
\begin{equation}
\dot P_i = -(1/4\pi)
\oint l_i |\lambda^0|^2 dS.
\end{equation}

The angular momentum is given by
\begin{equation}
M_A=-(1/4\pi)\Re \oint \bar l_A {\cal K}dS,
\label{AM}
\end{equation}
where $l_A$ is the solution of $\eth l_A = 0$, and
\begin{eqnarray}
{\cal K}=2\Psi^0_1 + 3\eth\eta(2\Psi^0_2 + 2\sigma^0{\cal N}- \bar\Psi^0_2
- \bar\sigma^0 \bar{\cal N}) \nonumber \\
+ (\eta - u)\left(2\eth\Psi^0_2 +
2\eth(\sigma^0{\cal N})- \eth\bar\Psi^0_2
- \eth(\bar\sigma^0 \bar{\cal N})\right ).  
\end{eqnarray}
The loss of  angular momentum is found to be
\begin{eqnarray}
\dot M_A = -(1/4\pi)\Re \oint \bar l_A {\cal G} dS,
\end{eqnarray}
where we set ${\cal G}= V^3\partial/\partial u({\cal K}/V^3) $.

Let $\Sigma_1$ and $\Sigma_2$ be arbitrary cross-sections at $\cal J^+$, then
the flux of the energy-momentum and angular momentum is given by
\begin{eqnarray}
P_i(\Sigma_2) - P_i(\Sigma_1) = -(1/4\pi) \oint B_i |{\cal N}|^2 dS du,\\
M_A(\Sigma_2) - M_A(\Sigma_1) = -(1/4\pi)\Re \oint \bar l_A {\cal G}
dS du,
\end{eqnarray}
where integration is performed over the domain $\Omega \subset {\cal J^+}$
contained between $\Sigma_1$ and $\Sigma_2$.

\section{Concluding Remarks}

Note that the group of isometries can be defined also as a group which
transforms an arbitrary geodesic to a geodesic one and the Killing vectors
satisfy the geodesic deviation equation for any geodesic.  In our
construction above (Sec. 3) only {\em null} geodesics passing inward are
transformed to the geodesics under the transformations of the Poincar\'e
quasigroup.  For non-radiating at $\cal J^+$ systems the generators
the Poincar\'e quasigroup is isomorphic to the Poincar\'e group. It agrees
with the well known results on the reduction of the BMS group to the
Poincar\'e group for the asymptotically flat stationary spacetime.

Our definition of the Noether charge associated with an arbitrary elements
of the Poincar\'e quasialgebra is free from the supertranslational
ambiguity and yields identically vanishing  charge/flux in a flat
spacetime.  It essentially depends on the geometry of future null infinity
and the behavior of generators near of $\cal J^+ $. The integral conserved
quantities $P(\xi, \Sigma) $  and a flux integral, giving the difference
$P(\xi,\Sigma') - P(\xi,\Sigma)$  for the cross-sections $\Sigma'$ and
$\Sigma$, are linear on generators of Poincar\'e quasigroup and defined
for system with radiation on  ${\cal J}^+$.

\section*{Acknowledgments}

This work was supported by CONACyT, Grant No. 1626P-E.

\end{document}